\documentclass[%
 reprint,
 amsmath,amssymb,
 aps,
 prd
floatfix,
showkeys
]{revtex4-1}

\usepackage[USenglish]{babel}
\usepackage{slashed}
\usepackage{graphicx}
\usepackage{dcolumn}
\usepackage{bm}

\begin{document}

\title{Magnetic field-dependence of the neutral pion mass in the linear sigma model coupled to quarks: The weak field case}


\author{Alejandro Ayala$^{1,2}$, Ricardo L. S. Farias$^3$, S. Hern\'andez-Ortiz$^1$, L. A. Hern\'andez$^{1,2}$, D. Manreza Paret$^{1,4}$ and R. Zamora$^{5,6}$}
\affiliation{%
$^1$Instituto de Ciencias Nucleares, Universidad Nacional Aut\'onoma de M\'exico, Apartado Postal 70-543, CdMx 04510, Mexico.\\
$^2$Centre for Theoretical and Mathematical Physics, and Department of Physics, University of Cape Town, Rondebosch 7700, South Africa.\\
$^3$Departamento de Fisica, Universidade Federal de Santa Maria, Santa Maria, RS 97105-900, Brazil.\\
$^4$Facultad de F\'isica, Universidad de La Habana, San Lazaro y L, La Habana, Cuba.\\
$^5$Instituto de Ciencias B\'asicas, Universidad Diego Portales, Casilla 298-V, Santiago, Chile.\\
$^6$Centro de Investigaci\'on y Desarrollo en Ciencias Aeroespaciales (CIDCA), Fuerza A\'erea de Chile,  Santiago, Chile.
}%


\begin{abstract}
We compute the neutral pion mass dependence on a magnetic field in the weak field approximation at one-loop order. The calculation is carried out within the linear sigma model coupled to quarks and using Schwinger's proper-time representation for the charged particle propagators. We find that the neutral pion mass decreases with the field strength provided the boson self-coupling magnetic field corrections are also included. The calculation should be regarded as the setting of the trend for the neutral pion mass as the magnetic field is turned on. 
\end{abstract}


\maketitle

\section{Introduction}\label{sec1}

Magnetic fields are involved in the properties of a large variety of physical systems including heavy-ion collisions~\cite{2008NuPhA.803..227K, MCLERRAN2014184}, the interior of compact astrophysical objects~\cite{Duncan:1992hi,2018arXiv180305716E,Ayala:2018kie} and even the early universe~\cite{VACHASPATI1991258,Navarro:2010eu,Sanchez:2006tt}. It has been estimated that the magnetic field strength $|eB|$ in peripheral heavy-ion collisions reaches values equivalent to a few times the pion mass squared, both at RHIC and at the LHC~\cite{2009IJMPA..24.5925S}. The effects of such magnetic fields cannot be overlooked in a complete description of these systems and its understanding contributes, at a fundamental level, to a better characterization of the properties of QCD matter~\cite{PhysRevC.91.064902, PhysRevD.90.085011, PhysRevD.92.096011, PhysRevD.92.016006, PhysRevD.91.016007,PhysRevD.91.016002, PhysRevD.89.116017, PhysRevD.89.016004, Ayala:2018wux}.   

One of these properties is the behavior of meson masses as a function of $|eB|$. These properties have been the subject of intense study in several recent works. Within the framework of Lattice QCD (LQCD) calculations, it has been shown that the neutral pion mass decreases monotonically as $|eB|$ grows~\cite{PhysRevD.97.034505}. The latter calculation solved an existing disagreement between the quenched Wilson and overlap fermions formulations of LQCD~\cite{ LUSCHEVSKAYA2015627,PhysRevD.87.094502}. 

In effective QCD models, the neutral pion mass has also been found to decrease with the increase of the magnetic field intensity. Some of these calculations resort to use the Nambu--Jona-Lasino (NJL) model and its extensions~\cite{2012PhRvD..86h5042F, 2018PhLB..782..155C, 2018PhRvD..97c4025G, 2016PhRvD..93a4010A}. In particular, Ref.~\cite{AVANCINI2017247} considers a two-flavor NJL model, employing a magnetic field dependent coupling, fitted to reproduce lattice QCD results~\cite{Bali:2012zg} for the quark condensates~\cite{2017EPJA...53..101F}. The $B$-dependence of the pion mass has also been studied using chiral perturbation theory~\cite{2001JHEP...10..006A, 2012PhRvD..86b5020A}. Also, Ref.~\cite{PhysRevD.93.074033} resorts to the use of phenomenological Lagrangians for the one-loop calculation of the pion effective mass in the weak field approximation $(|eB|<< m_\pi^2)$, finding that for a pseudoscalar coupling, the pion mass decreaces, whereas for a pseudovector coupling the mass increases, as a function of the magnetic field.    

In this work we explore the idea that magnetic field-dependent couplings, computed self-consistently, can account for the decrease of the neutral pion mass as a function of the magnetic field strength. For this purpose, we resort to use the Linear Sigma Model coupled to quarks (LSMq) to compute the neutral pion self-energy in the presence of a weak magnetic field, accounting for the one-loop corrections to the boson self-coupling. 
From the self-energy, we then obtain the magnetic field dependence of the neutral pion mass. The work is organized as follows. For completeness, in Sec.~\ref{sec2} we summarize the properties of the LSMq Lagrangian after both, spontaneous and explicit breaking of symmetry are implemented. In Sec.~\ref{sec3} the one-loop contribution to the pion self-energy in a weak magnetic field is computed taking into account the quark and meson contributions. In Sec.~\ref{sec4} we compute the magnetic field effects on the boson self-coupling. In Sec.~\ref{sec5} we compute the magnetic field effects on the vacuum expectation value. Finally, in Sec.~\ref{sec6} we find the explicit expression for the magnetic field-dependence of the neutral pion mass. We also discuss our results and conclude. We leave for the appendices the explicit calculations of the various quantities involved.

\section{Linear Sigma Model coupled to quarks with an explicit symmetry breaking term}\label{sec2}

The Lagrangian for the LSMq is given by
\begin{align}
  \mathcal{L}&=\frac{1}{2}(\partial_\mu \sigma)^2+\frac{1}{2}
  (D_\mu\vec{\pi})^2+\frac{a^2}{2}(\sigma^2+\vec{\pi}^2) \nonumber \\
  &-\frac{\lambda}{4}(\sigma^2+\vec{\pi}^2)^2+i\bar{\psi}\gamma^\mu D_\mu
  \psi-g\bar{\psi}(\sigma+i\gamma_5\vec{\tau}\cdot\vec{\pi})\psi,
\end{align}
where $\psi$ is an $SU(2)$ isospin doublet, $\vec{\pi}=(\pi_1,\pi_2,\pi_3)$ is an isospin triplet, and $\sigma$ is an isospin singlet, with
\begin{equation}
 D_\mu=\partial_\mu+iqA_\mu
\end{equation}
the covariant derivative. $A^\mu$ is the vector potential corresponding to an external magnetic field directed along the  $\hat{z}$ axis. In the symmetric gauge, it is given by
\begin{equation}
 A^\mu=\frac{B}{2}(0,-y,x,0).
\end{equation}
The gauge field couples only to the charged pion combinations, namely,
\begin{equation}
 \pi_\pm=\frac{1}{\sqrt{2}}(\pi_1\pm i\pi_2)
\end{equation}

To allow for spontaneous symmetry breaking, we let the $\sigma$ field to develop a vacuum expectation value $v$
\begin{equation}
 \sigma \rightarrow \sigma + v.
\end{equation}
After this shift, the Lagrangian can be rewritten as
\begin{align}
   \mathcal{L}&= \bar{\psi}(i\gamma^\mu D_\mu-M_f)\psi-g\bar{\psi}
   (\sigma+i\gamma_5\vec{\tau}\cdot \vec{\pi})\psi 
   - \frac{1}{2}M_\sigma^2\sigma^2\nonumber\\
   & - \frac{1}{2}M_\pi^2(\vec{\pi})^2
   -\frac{1}{2}(\partial_\mu\sigma)^2-
   \frac{1}{2}[(\partial_\mu+iqA_\mu)\vec{\pi}]^2\nonumber\\
   &-\lambda v(\sigma^2+\sigma\vec{\pi}^2) -\frac{1}{4}\lambda (\sigma^4+2
   \sigma^2\vec{\pi}^2+\vec{\pi}^4)\nonumber\\
   & +\frac{a^2}{2}v^2-\frac{\lambda}{4}v^4.
 \label{lagshift}
\end{align}
The quarks, sigma and the three pions have masses given by
\begin{align}
 M_f&=gv, \nonumber \\
 M_\sigma^2&=3\lambda v^2-a^2, \nonumber \\
 M_\pi^2&=\lambda v^2-a^2,
 \label{masses}
\end{align}
respectively. Also, we notice from Eq.~(\ref{lagshift}) that the tree-level potential is given by
\begin{equation}
 V^{\text{tree}}(v)=-\frac{a^2}{2}v^2+\frac{\lambda}{4}v^4,
\end{equation}
which has a minimum located at
\begin{equation}
 v_0=\sqrt{\frac{a^2}{\lambda}}.
\end{equation}
Therefore, the masses evaluated at $v_0$ are
\begin{align}
 M_f(v_0)&=g\sqrt{\frac{a^2}{\lambda}}, \nonumber \\
 M_\sigma^2(v_0)&=2a^2, \nonumber \\
 M_\pi^2(v_0)&=0.
\end{align}
In order to consider a non-vanishing pion mass, we add to the Lagrangian an explicit symmetry breaking term and thus
\begin{equation}
 \mathcal{L}\to\mathcal{L}'=\mathcal{L}+\frac{m_\pi^2}{2} v(\sigma+v),
\end{equation}
where $m_\pi\approx 140$ MeV. As a consequence of the explicit symmetry breaking, the $V^{\text{tree}}(v)$ is modified and therefore the minimum of the tree-level potential becomes
\begin{equation}
 v_0\to v_0^\prime=\sqrt{\frac{a^2+m_\pi^2}{\lambda}}.
\end{equation}
Consequently, although the expressions for the masses as functions of $v$, Eq.~(\ref{masses}), remain the same, their values at the minimum of the potential with explicit symmetry breaking become
\begin{align} \label{masses2}
 M_f(v_0) \to M_f(v_0^\prime)&=g\sqrt{\frac{a^2+m_\pi^2}{\lambda}}, \nonumber \\
 M_\sigma^2(v_0)\to M_\sigma^2(v_0^\prime)&=2a^2+3m_\pi^2, \nonumber \\
 M_\pi^2(v_0)\to M_\pi^2(v_0^\prime)&=m_\pi^2.
\end{align}
Also, notice that from Eq.~(\ref{masses}), we can fix the value of $a$ from the relation
\begin{eqnarray}
a&=&\sqrt{\frac{M_\sigma^2(v_0^\prime) - 3M_\pi^2(v_0^\prime)}{2}}\nonumber\\
&=&\sqrt{\frac{m_\sigma^2 - 3m_\pi^2}{2}}.
\label{relation}
\end{eqnarray}
For $m_\pi\approx 140$ MeV and when considering the sigma mass in the range $m_\sigma \approx 400 - 600$ MeV, we get $a\approx 225 - 390$ MeV. In the spirit of LQCD calculations, we later on consider also different values for $m_\pi$ and accordingly, we also extend the ranges for $m_\sigma$ and $a$.  

\section {One-loop pion self-energy}\label{sec3}

The self-energy for the neutral pion has four contributions
\begin{equation}
 \Pi(B,q)=\Pi_{f\bar{f}}(B,q)+\Pi_{\pi^\pm}(B)+\Pi_{\pi^0}+\Pi_{\sigma}.
\end{equation}
The one-loop diagrams for the first two contributions are depicted in Figs.~\ref{fig1} and~\ref{fig2}. Notice that for $\Pi_{\pi^0,\sigma}$ there are no magnetic corrections, given that the particles in the loop are neutral. We now proceed to compute the first two contributions to the self-energy.

\subsection{Quark loop}

\begin{figure}[t!]
 \begin{center}
  \includegraphics[scale=0.3]{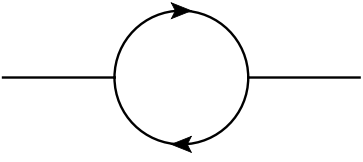}
 \end{center}
 \caption{One-loop Feynman diagram representing the quark-antiquark contribution to the $\pi^0$ self-energy.}
 \label{fig1}
\end{figure}

\begin{figure}[b!]
 \begin{center}
  \includegraphics[scale=0.14]{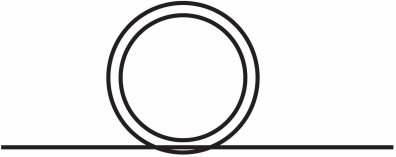}
 \end{center}
 \caption{One-loop Feynman diagram representing the charged pion contribution to the 
 $\pi^0$ self-energy.}
 \label{fig2}
\end{figure}

The contribution from the one-loop Feynman diagram made out of fermions to the neutral pion self-energy iss depicted in Fig.~\ref{fig1}.
In the presence of a magnetic field, this corresponds to the expression

\begin{equation}\label{propS1}
-i \Pi_{f\bar{f}}(B,q)=N_f g^2\int\frac{d^4k}{(2\pi)^4}\text{Tr}\{\gamma^5iS^B_f(k)\gamma^5iS^B_f(k-q)\},
\end{equation}
where $S^B_f(k)$ is the Schwinger's proper time representation of the charged fermion propagator.
In the weak magnetic field limit, this propagator can be written as a power series in $|eB|$ which, up to order $\mathcal{O}(eB)^2$, is written as~\cite{PhysRevD.62.105014}
\begin{align}\label{weakfieldprop}
 iS^B_f&(k)=i\frac{M_f+\slashed{k}}{k^2-M_f^2}-|q_fB|\frac{\gamma_1\gamma_2(M_f+\slashed{k}_\parallel)}{(k^2-M_f^2)^2}\nonumber \\
 &-2i(q_fB)^2\frac{k_\perp^2(M_f+\slashed{k}_\parallel)+\slashed{k}_\perp(M_f^2-k_\parallel^2)}{(k^2-M_f^2)^4},
\end{align}
where the transverse and parallel components of any vector are split, with respect to the magnetic field direction, and we use the definitions
\begin{align}
	\left(a\cdot b\right)_\parallel &= a^0 b^0 - a^3 b^3, \nonumber \\
    \left(a\cdot b\right)_\perp &= a^1 b^1 + a^2 b^2. 
\end{align}
Therefore, the contribution from the quark-antiquark loop to the self-energy can in turn be split into tree terms
\begin{equation}\label{eq2}
   \Pi_{f\bar{f}}(B,q)  = \Pi_{f\bar{f}}^0(q) + \Pi_{f\bar{f}} ^{|q_fB|}(q)+ \Pi_{f\bar{f}} ^{(q_fB)^2}(q).
\end{equation}
The first term is the usual vacuum piece which contributes only to the pion mass renormalization. The term linear in $|q_fB|$ gives a vanishing contribution.
The result for ${\mathcal{O}}(q_fB)^2$, in the limit of vanishing four-momentum $q\rightarrow 0$, is given by (see details in Appendix~\ref{ap1})
\begin{align}\label{eqsigma555}
  -i \Pi_{f\bar{f}} ^{(q_fB)^2}(q=0)  
 =-i\frac{N_f g^2(q_fB)^2}{4\pi^2M_f^2} .
\end{align}

\subsection{Meson loop}

The meson loops consist of tadpole diagrams, like the one depicted in Fig.~\ref{fig2}. The expression for the only diagram that contributes to the magnetic field correction to the pion mass is given by
\begin{align}\label{tad1}
  -i \Pi_{\pi^\pm} (B)  &= \frac{\lambda}{4} \int\frac{d^4k}{(2\pi)^4} D^B(k), 
\end{align}
where $D^B(k)$ is the propagator for a charged scalar boson in a magnetic field. In the weak field limit, this can also be expressed as a power series in $|eB|$ which, up to order $\mathcal{O}(eB)^2$, is  given by \cite{PhysRevD.71.023004}
\begin{align}\label{tad11}  
  iD^B(k)&=\frac{i}{k^2-M_\pi^2}
  -\frac{i}{[k^2-M_\pi^2]^3}(eB)^2 \nonumber \\
  &-\frac{2ik_\perp^2}{[k^2-M_\pi^2]^4}(eB)^2   .
\end{align}
The first term in Eq.~(\ref{tad11}) corresponds to the vacuum contribution, which is magnetic field independent and ultraviolet divergent. As usual, this term contributes to the pion mass renormalization. The magnetic field-dependent terms are finite. Thus, the one-loop contribution to the magnetic field corrections to the neutral pion mass, up to ${\mathcal{O}}(eB)^2$ due to charged mesons, is written as (see details in Appendix~\ref{ap2})
\begin{align}\label{tad5}
-i \Pi_{\pi^\pm} (B)  &=i\frac{\lambda}{4} \frac{(eB)^2}{96\pi^2}\left(\frac{1}{M_\pi^2}\right).
\end{align}

\subsection{One-loop magnetic modification to neutral pion mass}

In order to compute the magnetic field-induced modification to the pion mass, $M_\pi(B)$, we need to find the solution to the equation
\begin{eqnarray}
  q_0^2-|\vec{q}|^2-M_\pi^2-{\mbox{Re}}[\Pi(B,q)]=0
\label{findsol}
\end{eqnarray}
in the limit $\vec{q}\to 0$ and $q_0=M_\pi(B)$. 
For the time being, for simplicity, instead of working with the solution of Eq.~(\ref{findsol}), which requires a numerical treatment, let us first use the approximation
\begin{eqnarray}
{\mbox{Re}}[\Pi(B,q_0=M_\pi(B),\vec{q}=0)]\simeq \Pi(B,q_0=0,\vec{q}=0),
\nonumber\\
\label{approximate}
\end{eqnarray}
which allows for an analytic treatment. Nevertheless, in Sec.~\ref{sec6}, we show the results when explicitly solving Eq.~(\ref{findsol}).
Using Eqs.~(\ref{eqsigma555}) and (\ref{tad5}) we get
\begin{eqnarray}\label{modpi11}
\Pi(B,0)=\sum_f\frac{g^2(q_fB)^2}{4\pi^2M_f^2}-\frac{\lambda}{4} \frac{(eB)^2}{96\pi^2}\left(\frac{1}{M_\pi^2}\right),
\end{eqnarray}
where we sum over the number of quark-flavors $N_f$, and thus account for the absolute values of their electric charge. Using Eq.~(\ref{masses2}) and simplifying, we obtain
\begin{align}\label{modpi111}
 \Pi(B,0)&=\sum_f\frac{\lambda(q_fB)^2}{4\pi^2(a^2+m_\pi^2)}-\frac{\lambda}{4} \frac{(eB)^2}{96\pi^2m_\pi^2} \nonumber \\
 &=\frac{\lambda(eB)^2}{4\pi^2m_\pi^2} \left(\frac{5/9}{1+\frac{a^2}{m_\pi^2}}-\frac{1}{96}\right).
\end{align}
Notice that the dependence of the coupling $g$ cancels in the above expression.
Therefore, the magnetic field-dependent  neutral pion mass is given by
\begin{eqnarray}
   \!\!\!\!\!M_\pi^2(B)=\lambda {v'}_0^2-a^2+\frac{\lambda(eB)^2}{4\pi^2m_\pi^2} \left(\frac{5/9}{1+\frac{a^2}{m_\pi^2}}-\frac{1}{96}\right),
\label{modif}
\end{eqnarray}
where we recall that  $m_\pi$ is the vacuum pion mass. 
Since the parameters $a$ and $m_\pi$ are of the same order,  $5/9/(1+a^2/m_\pi^2)>1/96$. Naively it would seem that the neutral pion mass {\it increases} with the magnetic field strength. 
However, before drawing this conclusion, we observe that the calculation is not yet complete given that we also need to incorporate the one-loop magnetic-field correction to $\lambda$ and $v_0$. We now set up to find such corrections.

\section{Magnetic correction to the boson self-coupling}\label{sec4}

In order to compute the magnetic field correction to $\lambda$, we compute the Feynman diagrams depicted in Fig.~3.
\begin{figure}[ht!]
 \begin{center}
  \includegraphics[scale=0.6]{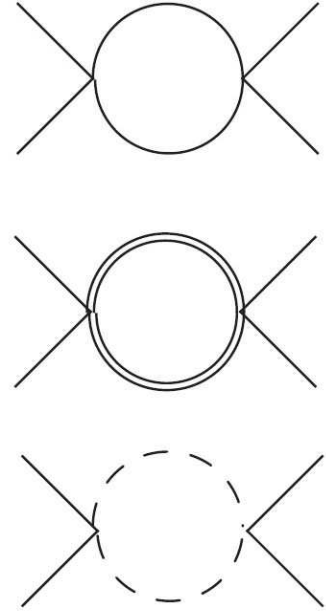}
 \end{center}
 \caption{One-loop Feynman diagrams contributing to the magnetic field correction to the self-coupling $\lambda$. The double line denotes the charged pion, the solid line is the neutral pion and the dashed line represents the sigma.}
 \label{fig3}
\end{figure}
We write the effective coupling up to one-loop order as 
\begin{equation}\label{deltalambda}
 \lambda^{\text{eff}}=\lambda+ \Delta \lambda,
\end{equation}
where $\Delta \lambda$ is given by~\cite{PhysRevD.91.016002},
\begin{align}\label{deltalambda0}
 \Delta \lambda&=24 \frac{\lambda^2}{16}[9I(q\rightarrow 0,M_\sigma^2)+I(q\rightarrow 0,M_\pi^2)\nonumber \\
 &+4J(q\rightarrow 0,M_\pi^2)],
\end{align}
where
\begin{align}
 I(q,M_i^2)  &=\int\frac{d^4k}{(2\pi)^4} D_i(q-k) D_i(k),
\end{align}
is the contribution from neutral boson fields in the loop ($i=\sigma,\pi^0$), with
\begin{align}
  D_i(k)  &= \frac{1}{k^2-M_i^2},
\end{align}
and
\begin{align}\label{Jj}
J(q,M_i^2) &=\int\frac{d^4k}{(2\pi)^4} D_i^B(q-k) D_i^B(k),
\end{align}
corresponds to the contribution from charged boson fields in the loop ($i=\pi^{\pm}$). The charged-boson propagator is given in Eq.~(\ref{tad11}). 
After computing both contributions and considering the pure magnetic correction, we obtain in the limiting of vanishing external four-momentum (see details in Appendix~\ref{ap3})
\begin{equation}
 \Delta \lambda=-\frac{27\lambda^2}{160\pi^2}\frac{(eB)^2}{M_\pi^4},
\end{equation}
and therefore, using Eqs.~(\ref{deltalambda}) and (\ref{masses2}) we have
\begin{equation}\label{deltalambda1}
 \lambda^{\text{eff}}=\lambda \Big[1-\frac{27\lambda}{160\pi^2}\frac{(eB)^2}{m_\pi^4}\Big].
 \end{equation}
 Notice that the magnetic field-induced corrections to the boson self-coupling make the effective coupling to {\it decrease} with the increase of the magnetic field strength.

\section{Magnetic correction to the vacuum expectation value}\label{sec5}
To include the magnetic correction to the vacuum expectation value, we start from the one-loop effective potential~\cite{PhysRevD.91.016002}. This includes the tree-level potential, the one-loop boson contribution, the one-loop fermion contribution and the counter-terms that come from requiring the vacuum stability conditions. These contributions are
\begin{figure}[t]
  \begin{center}
   \includegraphics[scale=0.45]{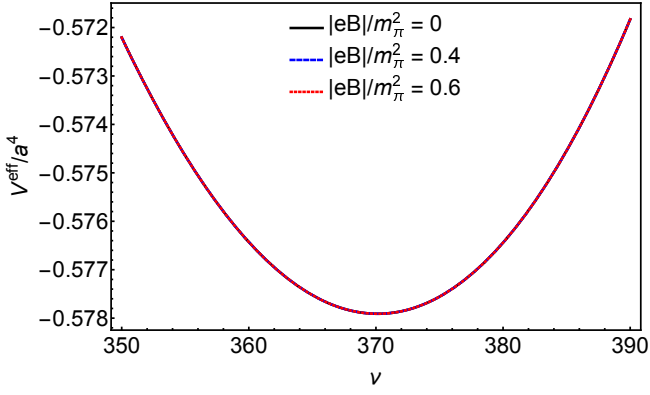}
  \end{center}
\caption{Magnetic field-dependent effective potential for different values of the field strength near the minimum $v_0^B$. Shown are three cases $|eB|/m_\pi^2=$ 0, 0.4 and 0.6, with $m_\pi=140 \ \text{MeV}$. Notice that the three curves are one on top of the others. Therefore, the minimum does not differ from $v_0'$.}
  \label{fig4}
 \end{figure}
\begin{align}\label{v01}
V^{cl} &=-\frac{(a^2+m_\pi^2)}{2}v^2+\frac{\lambda}{4}v^4, \\
V^{1}_b&=3\left\{\frac{(\lambda v^2-a^2)^2}{64\pi^2}\left[\ln\left(\frac{\lambda v^2-a^2}{a^2}\right)+\frac{1}{2}\right]\right\}\nonumber \\
&+\left\{\frac{(3\lambda v^2-a^2)^2}{64\pi^2}\left[\ln\left(\frac{3\lambda v^2-a^2}{a^2}\right)+\frac{1}{2}\right]\right\}\nonumber \\
&-2\left[\frac{(eB)^2}{192\pi^2}\ln\left(\frac{\lambda v^2-a^2}{a^2}\right)\right], \\
V^{1}_f&=-2\left\{\frac{(gv)^4}{16\pi^2}\left[\ln\left(\frac{(gv)^2}{a^2}\right)+\frac{1}{2}\right]\right\}\nonumber \\
&-\frac{5}{9}\left[\frac{(eB)^2}{24\pi^2}\ln\left(\frac{(gv)^2}{a^2}\right)\right],\\
V_c&=-\frac{\delta a^2}{2}v^2+\frac{\delta\lambda}{4}v^4,
\end{align}
respectively. The vacuum stability conditions~\cite{Carrington:1991hz} are introduced to ensure that $v_0$ and the sigma-mass maintain their tree level values, even after including the vacuum pieces stemming from the one-loop corrections. These conditions are
\begin{align}\label{v02}
\frac{1}{2v}\frac{dV^{\text{vac}}}{dv} \Big|_{v=v_0}&=0,\nonumber \\
\frac{d^2V^{\text{vac}}}{dv^2} \Big|_{v=v_0}&=2a^2+3m_\pi^2,
\end{align}
where $V^{\text{vac}}$ is the one-loop vacuum piece of the effective potential. The counterterms $\delta a^2$ and $\delta \lambda$ are given by 
\begin{eqnarray}
    \delta a^2&=&\frac{m_\pi^2}{2}
    -\frac{1}{16\pi^2 \lambda}\left\{\frac{}{} 3\lambda^2(6 a^2+4 m_\pi^2) -8 g^4(a^2+m_\pi^2) \right.\nonumber \\
    &+&\left.3 a^2 \lambda ^2 \Big[\ln  \Big(\frac{m_\pi^2}{a^2}\Big)+ \ln\Big(\frac{ 2a^2+3m_\pi^2}{a^2}\Big)\Big]\right\},
\end{eqnarray}
\begin{eqnarray}
\delta \lambda    &=&\frac{\lambda}{2}\left(\frac{ m_\pi^2}{a^2+ m_\pi^2}\right)\nonumber\\
    &-&\frac{1}{16\pi^2} \left\{\frac{}{}-16  g^4+24\lambda^2\right. -8 g^4 \ln  \left(g^2 \frac{(a^2+m_\pi^2)}{a^2\lambda}\right)\nonumber \\
    &+&\left. 3 \lambda ^2\left[ \ln  \left(\frac{m_\pi^2}{a^2}\right)
    +3 \ln  \left(\frac{ 2a^2+3m_\pi^2}{a^2}\right)\right]\right\}.
\end{eqnarray}
\begin{figure}[t!]
 \begin{center}
    \includegraphics[scale=0.48]{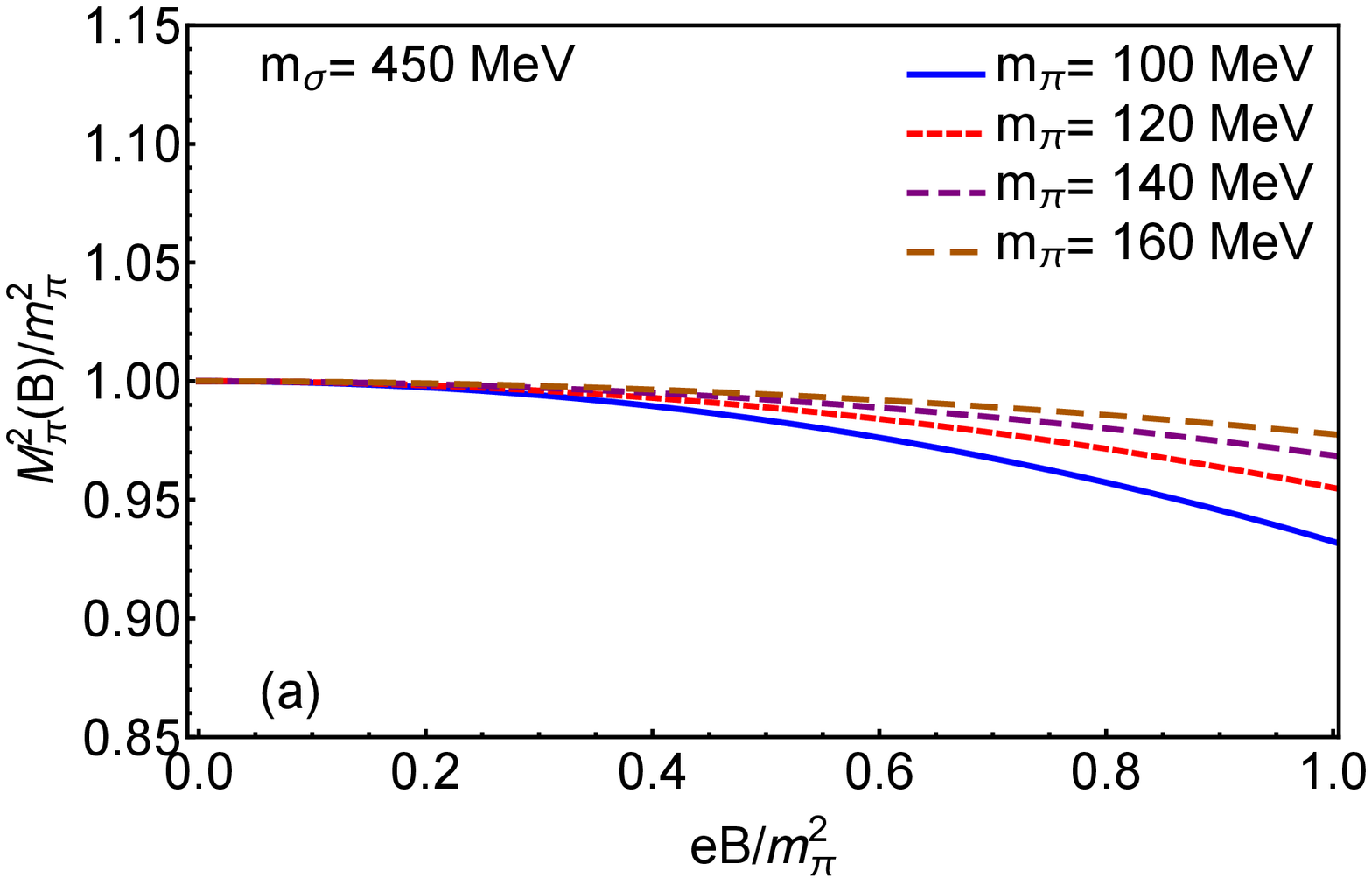}
    \includegraphics[scale=0.48]{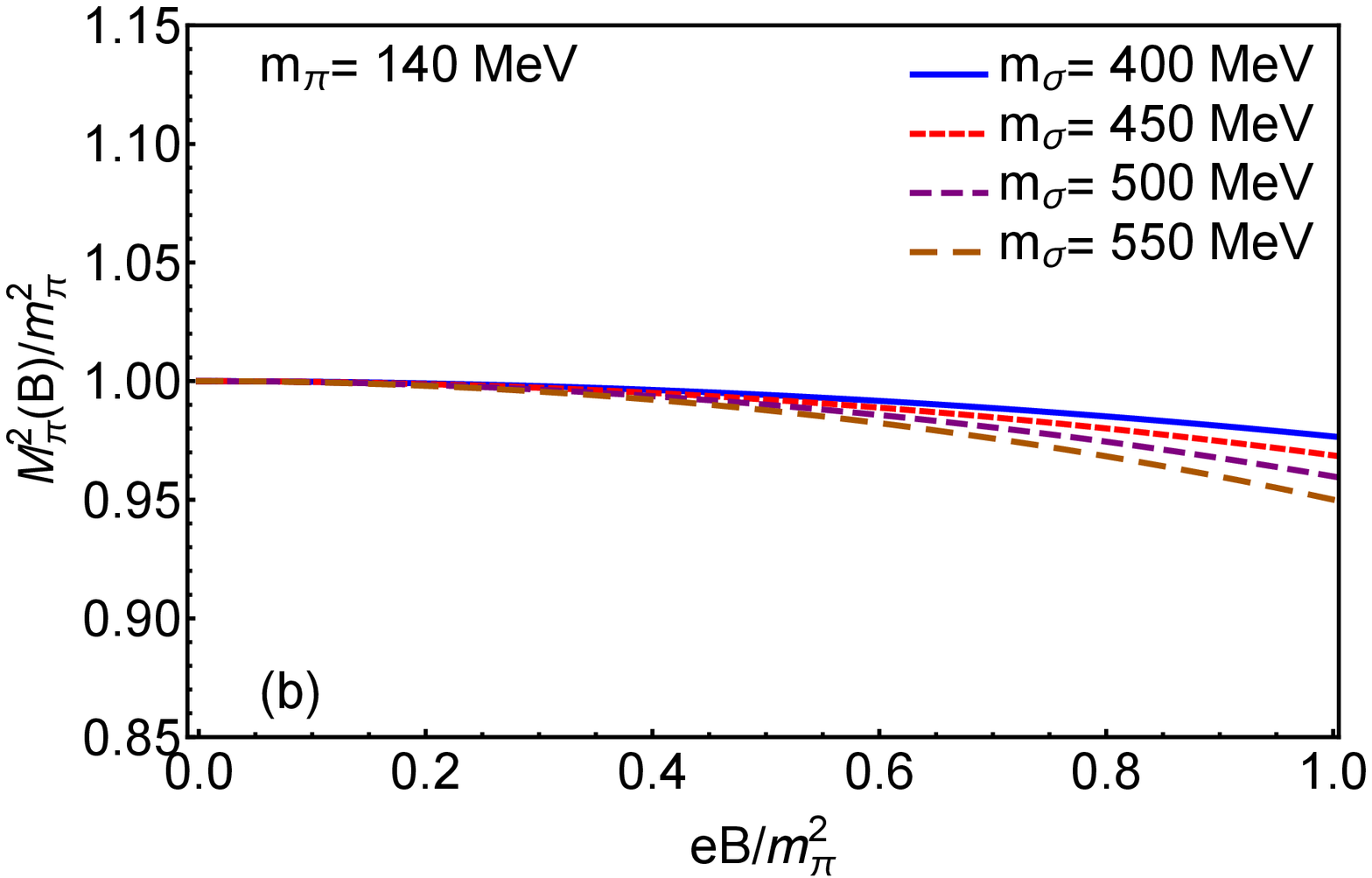}
 \end{center}
 \caption{Magnetic field dependence of the neutral pion mass (a) for a fixed vacuum $m_\sigma=450$ MeV, varying the vacuum pion mass between $m_\pi=100 - 160$ MeV, every 20 MeV and (b) for a fixed vacuum $m_\pi=140$ MeV varying the vacuum sigma mass between $m_\sigma=400 - 550$ MeV, every 50 MeV.}
 \label{fig5}
\end{figure}
The magnetic field-correction to the vacuum expectation value is obtained by finding the magnetic field displaced new minimum, $v_0^B$. It turns out that in the weak field limit, $v_0^B\simeq {v'}_0$, that is, the corrections are negligible. This is shown in Fig.~\ref{fig4}. 

\section{Discussion and conclusions}\label{sec6}

Since the corrections to the magnetic field-displaced position of the minimum are negligible, the significant correction to $M_\pi^2(B)$ in Eq.~(\ref{modif}) comes from the effective boson self-coupling and thus, the magnetic field-modified pion mass can be written as
\begin{eqnarray}\label{Mpi1}
 M_\pi^2(B)&\simeq&\lambda^{\text{eff}}{v'}_0^2-a^2+\frac{\lambda^{\text{eff}}(eB)^2}{4\pi^2m_\pi^2} \left(\frac{5/9}{1+\frac{a^2}{m_\pi^2}}-\frac{1}{96}\right).\nonumber \\
\end{eqnarray}
Substituting Eq.~(\ref{deltalambda1}) in  Eq.~(\ref{Mpi1}), simplifying and keeping terms up to ${\mathcal{O}}(eB)^2$, the magnetic field-modified pion mass is given by
\begin{eqnarray}\label{Mpi2}
 M_\pi^2(B)&\sim& m_\pi^2\left\{1-\frac{\lambda(eB)^2}{4\pi^2 m_\pi^4} \left[  \frac{27}{40}\Bigg(\frac{a^2}{m_\pi^2}+1\Bigg) \right.\right.\nonumber \\
&-&\left.\left. \left(\frac{5/9}{1+\frac{a^2}{m_\pi^2}}-\frac{1}{96}\right) \right]\right\}.
\end{eqnarray}

\begin{figure}[t!]
 \begin{center}
    \includegraphics[scale=0.48]{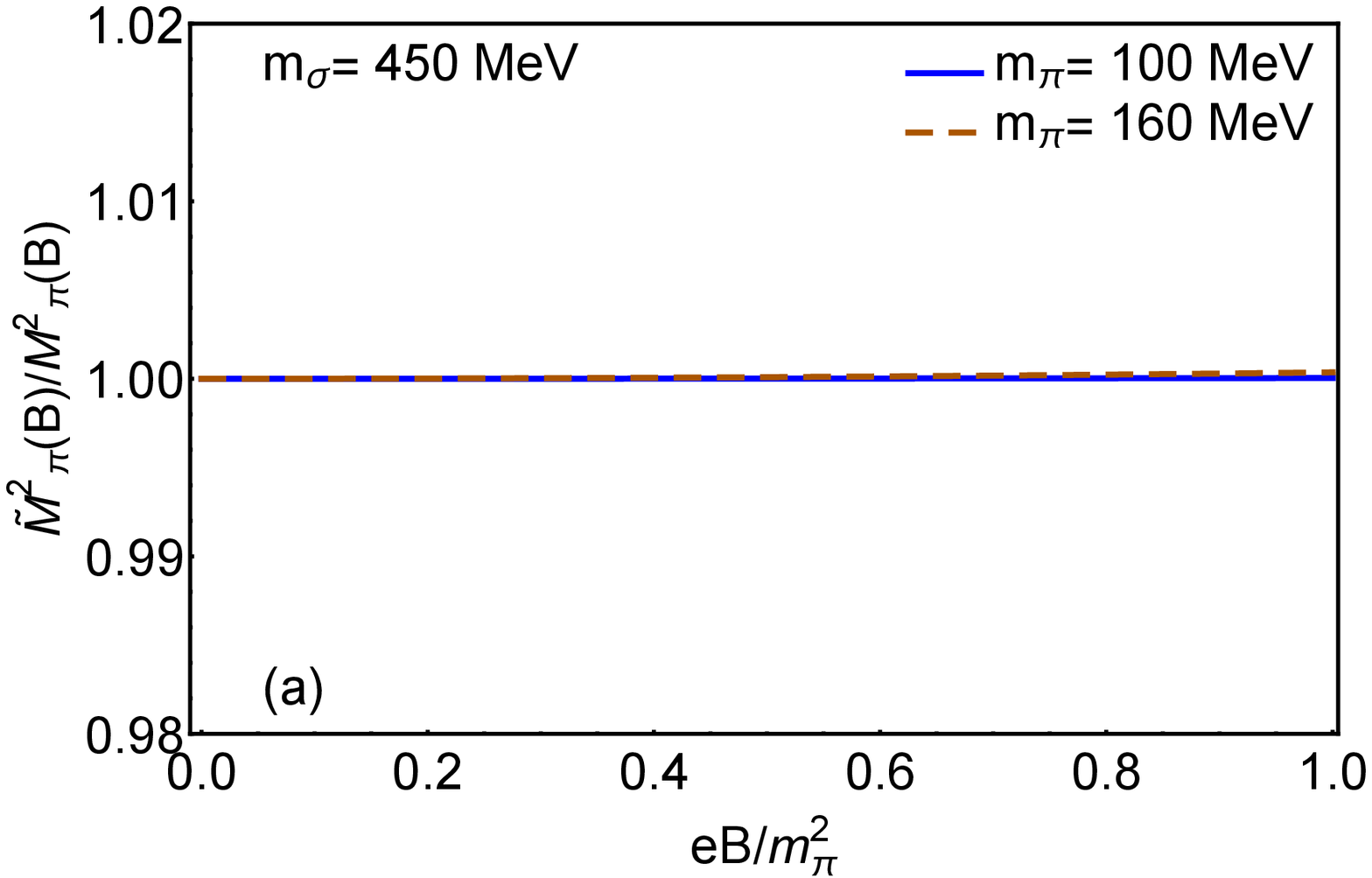}
    \includegraphics[scale=0.48]{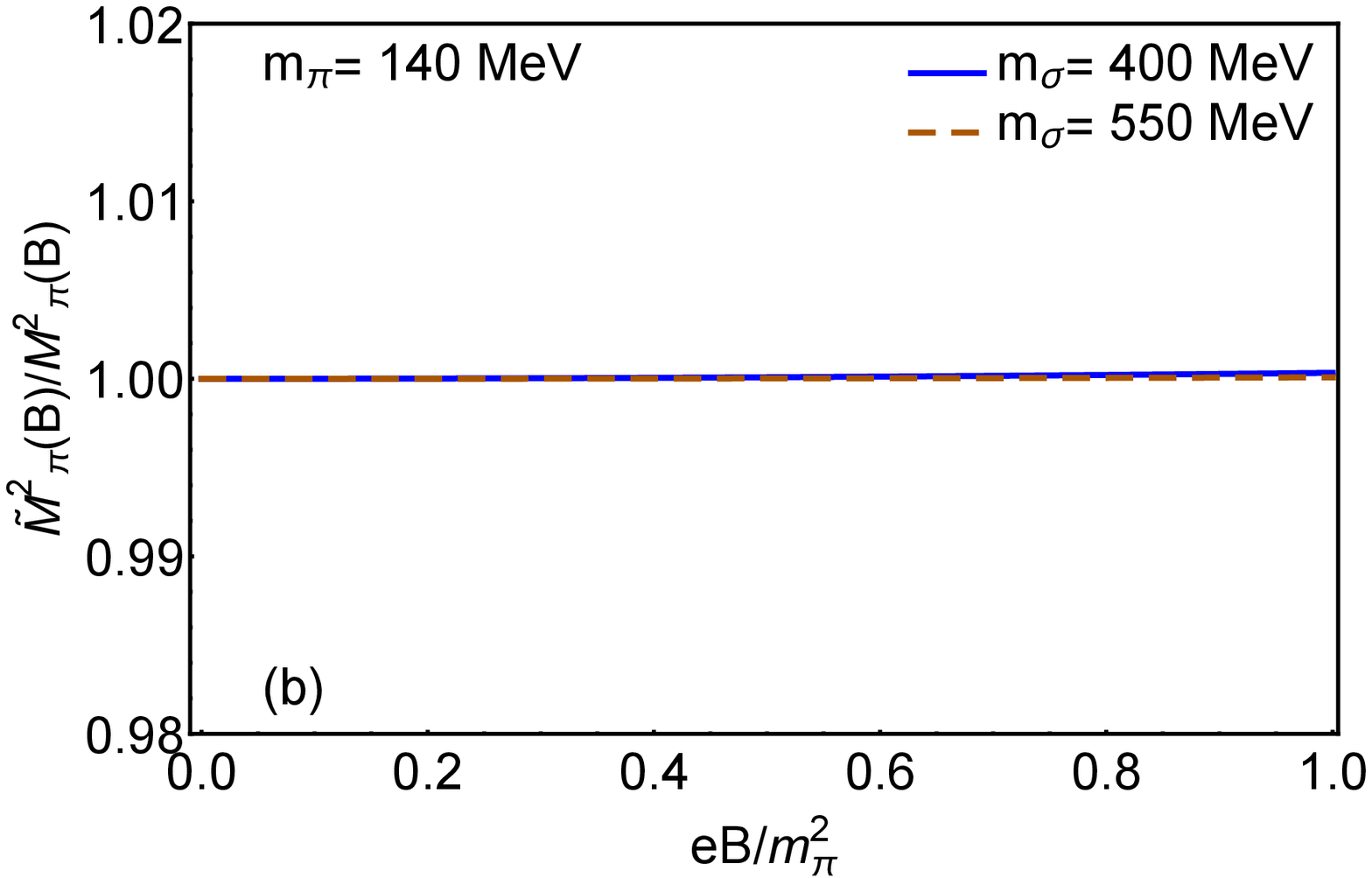}
 \end{center}
 \caption{Magnetic field dependence of the ratio of the neutral pion mass, when working with the numerical solution of Eq.~(\ref{findsol}) ($\tilde{M}^2_\pi$), to the case where we use the approximate solution given by Eq.~(\ref{approximate}) ($M_\pi^2$). (a) is the case for a fixed vacuum $m_\sigma=450$ MeV, for two vacuum pion massess $m_\pi=100,\ 160$ MeV and (b) is the case for a fixed vacuum $m_\pi=140$ MeV, for two vacuum sigma masses $m_\sigma=400,\ 550$ MeV.}
 \label{fig6}
\end{figure}
Figure~\ref{fig5} shows the neutral pion mass dependence on the  the magnetic field, obtained from Eq.~(\ref{Mpi2}) when varying the sigma and pion vacuum masses. Figure~\ref{fig6} shows the ratio of the neutral pion mass dependence on the magnetic field, when working with the numerical solution of Eq.~(\ref{findsol}) ($\tilde{M}^2_\pi$) to the case where we use the approximate solution ($M_\pi^2$), given by Eq.~(\ref{approximate}). Figure~\ref{fig6}a shows the case for a fixed vacuum value $m_\sigma=450$ MeV, for two values of the vacuum $m_\pi=100,\ 160$ MeV. Figure~\ref{fig6}b shows the case for a fixed vacuum value $m_\pi=140$ MeV, for two values of the vacuum $m_\sigma=400,\ 550$ MeV. Notice that using the exact and the approximate solutions of Eq.~(\ref{findsol}) does not make a difference. We stress that the calculation is limited to the weak field case and thus it has to be regarded as the trend for the neutral pion mass obtained as the magnetic field is turned on.

In conclusion, we have used the LSMq with spontaneous and explicit symmetry breaking to compute the magnetic field dependence of the neutral pion mass, including magnetic corrections to the one-loop pion self-energy, to the boson self-coupling and to the sigma field vacuum expectation value, in the weak field limit. Although the latter is a negligible correction, we found that the correction to the boson self-coupling produces that the pion mass {\it decreases} as a function of $|eB|$, which is opposite to the naive calculation obtained when ignoring self-coupling corrections. 

The results in this work should be regarded as the setting of the trend for magnetic field dependence of the neutral pion mass as the magnetic field is turned on. In order to compare this calculation to recent LQCD data, we require to extend the validity of the calculation to include the region of intermediate and large magnetic field compared to the vacuum pion mass. Work along these lines is currently being performed and will be reported elsewhere.

\section*{Acknowledgements}
Support for this work has been received in part by Consejo Nacional de Ciencia y Tecnolog\'ia grant number 256494, by Fondecyt (Chile) grant numbers 1170107, 1150471, 11508427 and Conicyt/PIA/Basal (Chile) grant number FB0821. Work partially supported from Conselho Nacional de Desenvolvimento Cient\'ifico e Tecnol\'ogico (CNPq), grant number 304758/2017-5 (R.L.S.F). R. Z. would like to acknowledge support from CONICYT FONDECYT Iniciaci\'on under grant number 11160234. D. M. acknowledges support from a PAPIIT-DGAPA-UNAM fellowship. 

\appendix

\section{Quark loop}\label{ap1}
To compute  the  $\mathcal{O}(eB)^2$ term that comes from substituting Eq.~(\ref{weakfieldprop}) in Eq.~(\ref{propS1}), we need to explicitly evaluate the expression 
\begin{widetext}
\begin{eqnarray}\label{Aeqsigma5}
  -i \Pi_{f\bar{f}} ^{(q_fB)^2}(q)  &=&N_f g^2(q_fB)^2\int\frac{d^2k_\parallel }{(2\pi)^2}\frac{d^2k_\perp}{(2\pi)^2} \left\{ \frac{\text{Tr}\{\gamma^5\gamma_1\gamma_2(M_f+\slashed{k}_\parallel)\gamma^5\gamma_1\gamma_2[M_f+(\slashed{k}_\parallel-\slashed{q}_\parallel)]\}}
  {[k^2-M_f^2]^2[(k-q)^2-M_f^2]^2}   \right. \nonumber \\
  &+& 4\left.\frac{\text{Tr}\{\gamma^5[k_\perp^2(M_f+\slashed{k}_\parallel)+\slashed{k}_\perp(M_f^2-k_\parallel^2)]\gamma^5[M_f+(\slashed{k}-\slashed{q})]\}}
  {[k^2-M_f^2]^4[(k-q)^2-M_f^2]} \right\}\nonumber\\
  &=&N_f g^2(q_fB)^2\int\frac{d^2k_\parallel }{(2\pi)^2}\frac{d^2k_\perp}{(2\pi)^2} 
   \left\{ \frac{\text{Tr}\{\gamma^5i(O^{+}-O^{-})(M_f+\slashed{k}_\parallel)\gamma^5i(O^{+}-O^{-})[M_f+(\slashed{k}_\parallel-\slashed{q}_\parallel)]\}}
  {[k^2-M_f^2]^2[(k-q)^2-M_f^2]^2}   \right. \nonumber \\
  &+& 4\left. \frac{\text{Tr}\{\gamma^5[k_\perp^2(M_f+\slashed{k}_\parallel)+\slashed{k}_\perp(M_f^2-k_\parallel^2)]\gamma^5[M_f+(\slashed{k}-\slashed{q})]\}}
  {[k^2-M_f^2]^4[(k-q)^2-M_f^2]} \right\} ,
\end{eqnarray}
\end{widetext}
where we introduced the spin-projection operators
\begin{equation}\label{Aopm}
  O^{\pm}=\frac{1}{2}(1\pm i\gamma_1\gamma_2) \,\, \Rightarrow \,\, \gamma_1\gamma_2=-i(O^{+}-O^{-}),
\end{equation}
that satisfy the identities
\begin{align}\label{Aopmid}
O^{\pm}O^{\pm}&=O^{\pm}, \,\, O^{\pm}O^{\mp}=0, \nonumber \\
 O^{\pm}\gamma_\perp^\mu&=\gamma_\perp^\mu O^{\mp}, \,\, O^{\pm}\gamma_\parallel^\mu=\gamma_\parallel^\mu O^{\pm}.
\end{align}
In order to compute the traces, we name the two terms in Eq.~(\ref{Aeqsigma5}) according to the power of $|q_fB|$ in each of the propagators of Eq.~(\ref{weakfieldprop}).
The first term comes from the product of the linear terms in $|q_fB|$
\begin{align}\label{Atr11}
\text{Tr}&\{\}_{11}= \text{Tr}\{\gamma^5iO^{+}(M_f+\slashed{k}_\parallel)\gamma^5iO^{+}[M_f+(\slashed{k}_\parallel-\slashed{q}_\parallel)]     \}\nonumber\\
+&\text{Tr}\{\gamma^5iO^{+}(M_f+\slashed{k}_\parallel)\gamma^5i(-O^{-})[M_f+(\slashed{k}_\parallel-\slashed{q}_\parallel)]     \}\nonumber\\
+&\text{Tr}\{\gamma^5i(-O^{-})(M_f+\slashed{k}_\parallel)\gamma^5iO^{+}[M_f+(\slashed{k}_\parallel-\slashed{q}_\parallel)]     \}\nonumber\\
+&\text{Tr}\{\gamma^5i(-O^{-})(M_f+\slashed{k}_\parallel)\gamma^5i(-O^{-})[M_f+(\slashed{k}_\parallel-\slashed{q}_\parallel)]   \},
\end{align}
using $(\gamma^5)^2=1$ and $\gamma^5\gamma^\nu=-\gamma^\nu\gamma^5$ we have
\begin{align}\label{Atr11_1}
\text{Tr}\{\}_{11}&= -\text{Tr}\{O^{+}(M_f-\slashed{k}_\parallel)O^{+}[M_f+(\slashed{k}_\parallel-\slashed{q}_\parallel)]     \}\nonumber\\
+&\text{Tr}\{O^{+}(M_f-\slashed{k}_\parallel)O^{-}[M_f+(\slashed{k}_\parallel-\slashed{q}_\parallel)]     \}\nonumber\\
+&\text{Tr}\{O^{-}(M_f-\slashed{k}_\parallel)O^{+}[M_f+(\slashed{k}_\parallel-\slashed{q}_\parallel)]     \}\nonumber\\
-&\text{Tr}\{O^{-}(M_f-\slashed{k}_\parallel)O^{-}[M_f+(\slashed{k}_\parallel-\slashed{q}_\parallel)]   \}.
\end{align}
Using the properties of $O^{\pm}$, we have
\begin{align}\label{Atr11_2}
\text{Tr}&\{\}_{11}= M_f^2\left( -\text{Tr}\{ O^{+}O^{+}\}+\text{Tr}\{ O^{+}O^{-}\}\right.\nonumber\\
&\left.+\text{Tr}\{ O^{-}O^{+}\}-\text{Tr}\{ O^{-}O^{-}\} \right)\nonumber\\
&+\text{Tr}\{ O^{+}\slashed{k}_\parallel O^{+}(\slashed{k}_\parallel-\slashed{q}_\parallel) \}-\text{Tr}\{ O^{+}\slashed{k}_\parallel O^{-}(\slashed{k}_\parallel-\slashed{q}_\parallel) \}\nonumber\\
&-\text{Tr}\{ O^{-}\slashed{k}_\parallel O^{+}(\slashed{k}_\parallel-\slashed{q}_\parallel) \}+\text{Tr}\{ O^{-}\slashed{k}_\parallel O^{-}(\slashed{k}_\parallel-\slashed{q}_\parallel) \},
\end{align}
and simplifying 
\begin{align}\label{Atr11_3}
\text{Tr}\{\}_{11}&=-M_f^2\text{Tr}\{ O^{+}+O^{-}\}\nonumber \\
&+\text{Tr}\{ (O^{+}+O^{-})\slashed{k}_\parallel(\slashed{k}_\parallel-\slashed{q}_\parallel) \}\nonumber \\
&=-4M_f^2+4k_\parallel\cdot(k_\parallel-q_\parallel).
\end{align}

The second term in Eq.~(\ref{Aeqsigma5})  comes from the product of the vacuum term times the quadratic term in $|q_fB|$
\begin{eqnarray}\label{Atr20}
\text{Tr}\{\}_{20}&=& 4\text{Tr} \{
\gamma^5[k_\perp^2(M_f+\slashed{k}_\parallel)\nonumber\\
&+&\slashed{k}_\perp(M_f^2-k_\parallel^2)]\gamma^5[M_f+(\slashed{k}-\slashed{q})]\}
\nonumber\\
&=&4k_\perp^2\text{Tr}\left\{\left[(M_f-\slashed{k}_\parallel)-\frac{\slashed{k}_\perp(M_f^2-k_\parallel^2)}{k_\perp^2}\right]\right.\nonumber\\
&\times& \left.[M_f+(\slashed{k}-\slashed{q})]\right\} \nonumber\\
&=&4k_\perp^2[4M_f^2-4k_\parallel\cdot(k-q)_\parallel \nonumber\\
&-&4\frac{(M_f^2-k_\parallel^2)}{k_\perp^2}k_\perp\cdot(k-q)_\perp].
\end{eqnarray}
Substitution of the traces in Eqs.~(\ref{Atr11_3}) and (\ref{Atr20}) into Eq.~(\ref{Aeqsigma5}), leads to
\begin{align}\label{Aeqsigma52}
  &-i \Pi_{f\bar{f}}^{(q_fB)^2}(q)=-4N_f g^2(q_fB)^2\int\frac{d^2k_\parallel }{(2\pi)^2}\frac{d^2k_\perp}{(2\pi)^2} \nonumber \\
  &\times \left\{ \frac{M_f^2-k_\parallel\cdot(k_\parallel-q_\parallel)}
  {[k^2-M_f^2]^2[(k-q)^2-M_f^2]^2}   \right. \nonumber \\
  & \left. - 4\frac{k_\perp^2[M_f^2-k_\parallel\cdot(k-q)_\parallel-\frac{(M_f^2-k_\parallel^2)}{k_\perp^2}k_\perp\cdot(k-q)_\perp]}
  {[k^2-M_f^2]^4[(k-q)^2-M_f^2]} \right\} .
\end{align}

We now introduce an integration over Feynman parameters. Equation~(\ref{Aeqsigma52}) contains two terms. We name them $K_1$ and $K_2$. These are given by~\cite{1995iqft.book.....P} 
\begin{align}\label{Afey_param2}
K_1&=\int_0^1dxdy\frac{4x^3\delta(x+y-1)}{[x(k^2-M_f^2)+y((k-q)^2-M_f^2)]^{5}} \nonumber\\
=&\int_0^1dx\frac{4x^3}{[x(k^2-M_f^2)+(1-x)((k-q)^2-M_f^2)]^{5}} \nonumber\\
=&4 \int_0^1dx\frac{x^3}{[l^2-\Delta]^{5}},
\end{align}
and
\begin{align}\label{Afey_param3}
K_2=&\int_0^1dxdy\frac{xy\delta(x+y-1)}{[x(k^2-M_f^2)+y((k-q)^2-M_f^2)]^{4}}\nonumber \\
\times&\frac{\Gamma(4)}{\Gamma(2)\Gamma(2)} \nonumber \\
=&6\int_0^1dx\frac{x(1-x)}{[x(k^2-M_f^2)+(1-x)((k-q)^2-M_f^2)]^{4}} \nonumber \\
=&6\int_0^1dx\frac{x(1-x)}{[l^2-\Delta]^{4}}.
\end{align}
where
\begin{align}\label{ADelta_l}
\Delta &=M_f^2-x(1-x)q^2, \\
l&= k-(1-x)q.
\end{align}
Substituting Eqs.~(\ref{Afey_param2}) and (\ref{Afey_param3}) into Eq.~(\ref{Aeqsigma52})
\begin{eqnarray}\label{Aeqsigma53}
-i \Pi_{f\bar{f}} ^{(q_fB)^2}(q)  &=&-4N_f g^2(q_fB)^2\int_0^1 dx\int\frac{d^2k_\parallel }{(2\pi)^2}\frac{d^2k_\perp}{(2\pi)^2}\nonumber\\
&\times&\left\{ \frac{6x(1-x)}{[l^2-\Delta]^{4}} [M_f^2-k_\parallel\cdot(k_\parallel-q_\parallel)]   \right. \nonumber \\
&-& 4\frac{4x^3}{[l^2-\Delta]^{5}} \left[k_\perp^2[M_f^2-k_\parallel\cdot(k-q)_\parallel]
\right.\nonumber\\
&-&\left.\left.(M_f^2-k_\parallel^2)k_\perp\cdot(k-q)_\perp \right] \right\} .
\end{eqnarray}
Setting $k_\parallel=l_\parallel+(1-x)q_\parallel$ and $k_\perp=l_\perp+(1-x)q_\perp$
\begin{widetext}
\begin{align}\label{Aeqsigma54}
-i \Pi_{f\bar{f}} ^{(q_fB)^2}(q)  &=-4N_f g^2(q_fB)^2\int_0^1 dx\int\frac{d^2l_\parallel }{(2\pi)^2}\frac{d^2l_\perp}{(2\pi)^2}  \left\{ \frac{6x(1-x)}{[l^2-\Delta]^{4}} [M_f^2-(l_\parallel+(1-x)q_\parallel)\cdot(l_\parallel-xq_\parallel)]  \right. \nonumber \\
& - 16\frac{x^3}{[l^2-\Delta]^{5}}  \left[(l_\perp+(1-x)q_\perp)^2[M_f^2-(l_\parallel+(1-x)q_\parallel)\cdot(l_\parallel-xq_\parallel)]  \right.  \nonumber \\
& \left. \left. -(l_\perp+(1-x)q_\perp)\cdot(l_\perp-xq_\perp)[M_f^2-(l_\parallel+(1-x)q_\parallel)^2]  \right]
\right\} .
\end{align}
\end{widetext}
Discarding odd powers of $l$ in Eq.~(\ref{Aeqsigma54}) we obtain 
\begin{widetext}
\begin{align}\label{Aeqsigma542}
  -i \Pi_{f\bar{f}} ^{(q_fB)^2}(q)&=-4N_f g^2(q_fB)^2\int_0^1 dx\int\frac{d^2l_\parallel }{(2\pi)^2}\frac{d^2l_\perp}{(2\pi)^2} \left\{ \frac{6x(1-x)}{[l^2-\Delta]^{4}} [M_f^2-l_\parallel^2+x(1-x)q_\parallel^2]  \right. \nonumber \\
  & - 16\frac{x^3}{[l^2-\Delta]^{5}}  \left[ -2l_\perp^2l_\parallel^2-(1-x)(1-2x)l_\parallel^2q_\perp^2  \right. l_\perp^2[2M_f^2-(1-x)(1-2x)q_\parallel^2]\nonumber \\
  & \left. \left. +(1-x)q_\perp^2[(1-2x)M_f^2+2x(1-x)^2q_\parallel^2]  \right]
\right\} .
\end{align}
\end{widetext}
To carry out the integration over $l_\parallel$ and $l_\perp$, we use the identity~\cite{Valenzuela:2014uia}
\begin{widetext}
\begin{align}\label{Aeqsigma55}
  i (-1)^{m-r}\int_0^1 dx \, x^\alpha (1-x)^\beta& \int\frac{d^2l_\parallel}{(2\pi)^2}\frac{d^2l_\perp}{(2\pi)^2}  \frac{(l_\perp^2)^n(l_\parallel^2)^m}{[l_\parallel^2+l_\perp^2+\Delta]^{r}}  \nonumber \\
  &=\frac{i (-1)^{m-r}}{16 \pi^2}B(n+1,r-n-1) B(m+1,r-n-m-2)\int_0^1 dx \, \frac{x^\alpha (1-x)^B}{\Delta^{r-n-m-2}},
\end{align}
\end{widetext}
where $B(m,n)$ is the Euler beta function.
Then, the $(q_fB)^2$ contribution to the $f\bar{f}$ loop becomes
\begin{widetext}
\begin{eqnarray}\label{Aeqsigma5512}
  -i \Pi_{f\bar{f}} ^{(q_fB)^2}(q)  &=&-4 i N_f g^2(q_fB)^2\int_0^1 dx\nonumber\\
  &\times&\left\{ \frac{6x(1-x)}{16\pi^2} \left[[M_f^2+x(1-x)q_\parallel^2] \frac{B(1,3)B(1,2)}{\Delta^2}+\frac{B(1,3)B(2,1)}{\Delta}\right]\right.\nonumber\\
  &-&16\frac{x^3}{16\pi^2}  \left[ -2\frac{B(2,3)B(2,1)}{\Delta}\right.-(1-x)(1-2x)q_\perp^2\frac{B(1,4)B(2,2)}{\Delta^2}\nonumber\\
  &-&[2M_f^2-(1-x)(1-2x)q_\parallel^2]\frac{B(2,3)B(1,2)}{\Delta^2} \nonumber\\
  &-&\left. \left. (1-x)q_\perp^2[(1-2x)M_f^2+2x(1-x)^2q_\parallel^2]
  \frac{B(1,4)B(1,3)}{\Delta^3}  \right]
\right\}.
\end{eqnarray}
\end{widetext}
Simplifying, we obtain
\begin{widetext}
\begin{align}\label{Aeqsigma552}
-i \Pi_{f\bar{f}} ^{(q_fB)^2}(q)  &=-\frac{4 i N_f g^2(q_fB)^2}{\pi^2}\int_0^1 dx  \left\{ \frac{x(1-x)}{16\Delta^2} \left[M_f^2+x(1-x)q_\parallel^2+\Delta\right]  \right.  - \frac{x^3}{24\Delta^3}  \left[ -2\Delta^2-(1-x)(1-2x)q_\perp^2\Delta  \right.  \nonumber \\
&-[2M_f^2-(1-x)(1-2x)q_\parallel^2]\Delta 
\left. \left. -2(1-x)q_\perp^2[(1-2x)M_f^2+2x(1-x)^2q_\parallel^2]  \right]
\right\} .
\end{align}
\end{widetext}
Further simplifications lead to
\begin{widetext}
\begin{align}\label{Aeqsigma553}
  -i \Pi_{f\bar{f}} ^{(q_fB)^2}(q)  &=-i \frac{N_f g^2(q_fB)^2}{4\pi^2}\int_0^1 dx \left\{ \frac{x(1-x)}{\Delta^2} \left[2M_f^2+x(1-x)q_\perp^2\right]  \right.  - \frac{2x^3}{3\Delta^3}  \left[ -2\Delta^2-(1-x)(1-2x)q_\perp^2\Delta  \right.\nonumber\\  &-[2M_f^2-(1-x)(1-2x)q_\parallel^2]\Delta\left. \left. -2(1-x)q_\perp^2[(1-2x)M_f^2+2x(1-x)^2q_\parallel^2]  \right]
\right\}.
\end{align}
\end{widetext}
Substituting the explicit expression for $\Delta$, we get
\begin{widetext}
\begin{align}\label{Aeqsigma554}
  &-i \Pi_{f\bar{f}} ^{(q_fB)^2} (q) =-i \frac{N_f g^2(q_fB)^2}{4\pi^2}\int_0^1 dx  \left\{ \frac{x(1-x)}{\Delta^2} \left[2M_f^2+x(1-x)q_\perp^2\right]  \right. \nonumber \\
  & \left. - \frac{2x^3}{3\Delta^3}  \left[-4M_f^4+M_f^2(1-x)[(1+4x)q_\parallel^2-3q_\perp^2]      -x(1-x)^2[(q_\parallel^2+q_\perp^2)^2-4xq_\parallel^2q_\perp^2]   \right]
\right\}.
\end{align}
\end{widetext}

Taking the limit $q^2\rightarrow0$ in Eq.~(\ref{Aeqsigma554}), we obtain
\begin{align}\label{Aeqsigma555}
  &-i \Pi_{f\bar{f}} ^{(q_fB)^2}(q)  =-i \frac{N_f g^2(q_fB)^2}{4\pi^2}\int_0^1 dx \nonumber \\
  &\times \left\{ \frac{x(1-x)}{M_f^4}2M_f^2  - \frac{2x^3}{3M_f^6} [-4M_f^4]\right\}  \nonumber \\
 &=-\frac{iN_f g^2(q_fB)^2}{4\pi^2}\int_0^1 dx\left\{ \frac{2x(1-x)}{M_f^2}  + \frac{8x^3}{3M_f^2}\right\}\nonumber \\
 &=-\frac{i N_f g^2(q_fB)^2}{4\pi^2M_f^2} .
\end{align}

\section{Meson loop}\label{ap2}
When the propagator of Eq.~(\ref{tad11}) is substituted into Eq.~(\ref{tad1}) we obtain the expression
\begin{align}\label{Btad1}
  -i \Pi_{\pi^\pm}(B)&= \frac{\lambda}{4} \int\frac{d^4k}{(2\pi)^4}\left\{ \frac{1}{k^2-M_\pi^2} \right. \nonumber \\
  & \left.-\frac{1}{[k^2-M_\pi^2]^3}(eB)^2-\frac{2k_\perp^2}{[k^2-M_\pi^2]^4}(eB)^2   \right\} .
\end{align}
The four-momentum integral is given by~\cite{1995iqft.book.....P}
\begin{align}\label{Btad3}
  -i \Pi_{\pi^\pm}(B)   &=\frac{\lambda}{4} \left\{ \frac{-i}{(4\pi^2)^{2-\epsilon}}\Gamma\left(\epsilon-1\right)\Big(\frac{1}{M_\pi^2}\Big)^{\epsilon-1}\right. \nonumber \\
  & + \frac{i}{2(4\pi^2)}\Big(\frac{1}{M_\pi^2}\Big)(eB)^2\nonumber \\
  &\left. - \frac{2 i}{16\pi^2}B(2,2)B(1,1)\frac{(eB)^2}{M_\pi^2}   \right\} .
\end{align}
where we employed dimensional regularization. Simplifying, we obtain
\begin{align}\label{Btad4}
  -i \Pi_{\pi^\pm}(B)   &=\frac{\lambda}{4} \left\{ -i\frac{M_\pi^2}{16\pi^2}\left[-\frac{1}{\epsilon}+\gamma_E-1-\ln(\frac{2\pi\mu^2}{M_\pi^2})\right]\right. \nonumber \\
  & \left.+ \frac{i(eB)^2}{96\pi^2}\left(\frac{1}{M_\pi^2}\right)\right\} .
\end{align}
Finally, using the $\overline{\text{MS}}$ scheme, with the ultraviolet scale $\mu$, the one-loop pure charged meson contribution is written as
\begin{align}\label{Btad5}
  -i \Pi_{\pi^\pm}(B)  &=\frac{\lambda}{4} \left\{ -i\frac{M_\pi^2}{16\pi^2}\left[-\ln\left(\frac{\mu^2}{M_\pi^2}\right)\right]\right. \nonumber \\
  & \left.+ \frac{i(eB)^2}{96\pi^2}\left(\frac{1}{M_\pi^2}\right)\right\},
\end{align}
where the first term comes from vacuum, and thus contributes to the renormalization pion mass. Accounting only for the magnetic contribution, we finally obtain
\begin{align}\label{Btad6}
-i \Pi_{\pi^\pm}&=\frac{\lambda}{4} \frac{i(eB)^2}{96\pi^2}\left(\frac{1}{M_\pi^2}\right).
\end{align}

\section{Magnetic correction to the boson self-coupling}\label{ap3}
We first compute $I(q,M_i^2)$. Its explicit form is
\begin{align}\label{Cl2}
 I(q,M_i^2)  &=\int\frac{d^4k}{(2\pi)^4} \frac{1}{[(q-k)^2-M_i^2]}\frac{1}{[k^2-M_i^2]}.
\end{align}
Using integration over Feynman parameters we get
\begin{align}\label{Cl22}
 I(q,M_i^2)  &=i\int_0^1 dx (2\pi)\left[\frac{1}{\epsilon}-\gamma_E+\ln(\frac{\Delta_i}{2\pi\mu^2})\right],
\end{align}
where $\Delta_i=M_i^2-x(1-x)q^2$. Notice that the above expression is magnetic field-independent. Therefore, $I(q,M_i^2)$ does not contribute to the $|eB|$ corrections to the boson self-coupling.

On the other hand, the explicit expression for $J(q,M_i^2)$, up to order $(eB)^2$ of Eq~(\ref{Jj}), is
\begin{align}\label{Cl3}
&J^{(eB)^2}(q,M_i^2)=\int\frac{d^4k}{(2\pi)^4}\Bigg \{ \Bigg [\frac{1}{[(q-k)^2-M_i^2]}\nonumber \\
  &-\frac{(eB)^2}{[(q-k)^2-M_i^2]^3}- \frac{2(eB)^2(q-k)_\perp}{[(q-k)^2-M_i^2]^4}\Bigg ]\nonumber \\
   & \times\Bigg [\frac{1}{[k^2-M_i^2]}-\frac{(eB)^2}{[k^2-M_i^2]^3}- \frac{2(eB)^2k_\perp^2}{[k^2-M_i^2]^4}\Bigg ]\Bigg \}.
\end{align}
After some simplifications, we obtain
\begin{align}\label{Cl4}
J^{(eB)^2}(q,M_i^2)  &=-2(eB)^2  \int\frac{d^4k}{(2\pi)^4}   \nonumber \\
   & \times\left\{ \frac{1}{[(q-k)^2-M_i^2][k^2-M_i^2]^3}\right.  \nonumber \\
   &+\left.\frac{2k_\perp^2}{[(q-k)^2-M_i^2][k^2-M_i^2]^4}\right\}.
\end{align}
Using integration over Feynman parameters and Eq.~(\ref{Aeqsigma55}), we arrive at
\begin{align}\label{Cl5}
J^{(eB)^2}(q,M_i^2)  &=-2(eB)^2\int_0^1  \nonumber \\
  &\times\left\{ \right. 3(1-x)^2\frac{(-1)^4i}{(4\pi)^2}\frac{\Gamma(2)}{\Gamma(4)}\left(\frac{1}{\Delta_i}\right)^2\nonumber \\
  &+8(1-x)^3\frac{(-1)^5i}{(4\pi)^2}B(2,3)B(1,2) \left(\frac{1}{\Delta_i}\right)^2 \nonumber \\
   &\left. +8(1-x)^3q_\perp^2\frac{(-1)^5i}{(4\pi)^2}\frac{\Gamma(3)}{\Gamma(5)}\left(\frac{1}{\Delta_i}\right)^3 \right\},
\end{align}
where $\Delta_i=M_i^2-x(1-x)q^2$. Simplifying
\begin{align}\label{Cl6}
J^{(eB)^2}&(q,M_i^2)=-i\frac{(eB)^2}{48\pi^2}\int_0^1 \frac{(1-x)^2}{\Delta_i^3} \nonumber \\
  &\times \left\{   [m^2-x(1-x)q^2](1+2x)-4(1-x)q_\perp^2 \right\},
\end{align}
taking the limit $q\rightarrow0$ and integrating
\begin{align}\label{Cl66}
J^{(eB)^2}(q,M_i^2)  &=-i\frac{9}{320}\frac{(eB)^2}{M_i^4}  .
\end{align}
Account for Eq.~(\ref{deltalambda0}) and the results from Eqs.~(\ref{Cl2}) and~(\ref{Cl66}), the correction to $\lambda$ can be written as
\begin{equation}\label{Cl8}
  \Delta\lambda=-\frac{27\lambda}{160\pi^2}\frac{(eB)^2}{M_i^4},
\end{equation}
where $i=\pi^{\pm}$.

%

\end{document}